# A simple method to characterize the electrical and mechanical properties of micro-fibers

A Castellanos-Gomez[*]

Kavli Institute of Nanoscience, Delft University of Technology, Lorentzweg 1, 2628 CJ Delft, The Netherlands.

E-mail: a.castellanosgomez@tudelft.nl

A procedure to characterize the electrical and mechanical properties of micro-fibers is presented here. As the required equipment can be found in many teaching laboratories, it can be carried out by physics and mechanical/electrical engineering students. The electrical resistivity, mass density and Young's modulus of carbon micro-fibers have been determined using this procedure, obtaining values in very good agreement with the reference values. The Young's modulus has been obtained by measuring the resonance frequency of carbon fiber based cantilevers. In this way, one can avoid common approaches based on tensile or bending tests which are difficult to implement for microscale materials. Despite the simplicity of the experiments proposed here, they can be used to trigger in the students interest on the electrical and mechanical properties of microscale materials.

### 1. Introduction

The characterization of the electrical and mechanical properties of novel materials is a milestone in the material science and mechanical and electronic engineering. While for macroscopic materials mechanical and electrical tests are well-defined [1-3], when the dimensions of the material are reduced down to the micron-scale, characterization of these properties become challenging [4, 5].

Here we present an experimental procedure to characterize the electrical and mechanical properties of carbon micro-fibers that can be carried out in a teaching laboratory by physics and mechanical/electrical engineering students. Although the methodology developed here can be applied to other micron-scale materials, we choose carbon fiber because of its emerging technological interest. For instance, carbon micro-fibers are regularly employed as fillers in composite materials to engineer their electrical and mechanical properties. Note that the characterization of the Young's modulus and electrical resistivity of individual micro-fibers is crucial to predict the properties of the composite material after adding the filler. Among different micro-fibers, we chose carbon fibers because of the availability in the market of many kinds of carbon fibers with different electronic and mechanical properties. Moreover, carbon fibers are very affordable and despite their small dimensions they can be handled with ease using normal laboratory tweezers. In the following sections the characterization of the electrical resistivity, mass density and Young's modulus are described.

## 2. Electrical characterization

A single carbon fiber is placed lying down straight onto a laboratory glass slide. Electrical contacts are made by means of small droplets of silver-loaded conductive epoxy adhesive[1]. In order to discount for the effect of the contact resistance ($R_C$) in the electrical measurements, four-terminal measurements are typically carried out. In this scheme, a current source is employed to inject current between two outer electrodes while a voltmeter is used to measure the voltage drop between two inner electrodes. An alternative and simpler approach consists on measuring the two-terminal resistance (directly with a multimeter) between contacts separated by an increasing distance. Figure 1 shows the measured two terminal resistance ($R$) as a function of the distance between electrical contacts ($L$). Both the contact resistance and the electrical resistivity ($\rho_{el}$) can be extracted from the $R$ *vs.* $L$ relationship:

$R = (\rho_{el} A^{-1}) \cdot L + R_C$,     [1]

Where $A$ is the carbon fiber cross section ($A = \pi r^2 = 38.5\ \mu m^2$). Therefore, fitting the measured data to a linear relationship one can determine the contact resistance from the interception with the vertical axis and the electrical resistivity from the slope. The value obtained for the electrical resistivity is $\rho_{el} = (1.3 \pm 0.3) \cdot 10^{-5}\ \Omega \cdot m$ while the contact resistance value is about 338 $\Omega$.

---

[1] Silver loaded epoxy. Purchased at RS-Online with part number 186-3616.

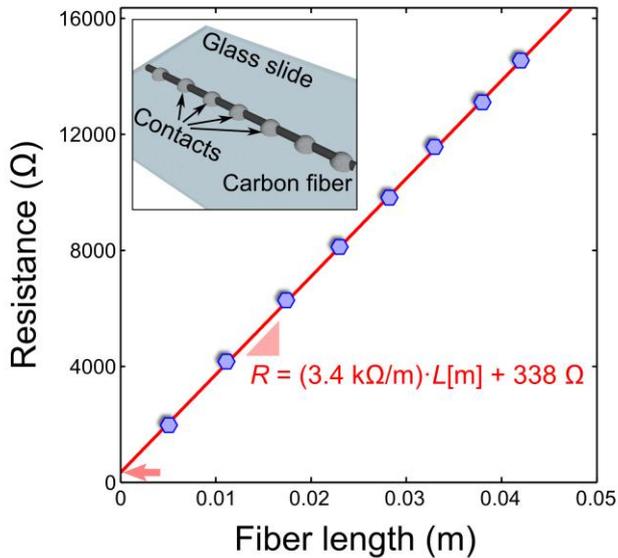

**Figure 1**: Two-terminal electrical resistance measured for increasingly separated electrical contacts. (Inset) schematic diagram of the carbon fiber with many electrical contacts made with silver-loaded conductive expoxy.

## 3. Mechanical characterization

### 3.1. Mass density

The mass density of carbon fibers can be determined by measuring the mass for a given material volume. Nonetheless, due to the combination of small dimensions of carbon fibers (typically 5 to 10 um in diameter) with their low values of mass density (ranging from 1.5 gr·cm$^{-3}$ to 2.5 gr·cm$^{-3}$), the mass of a single carbon fiber cannot be easily determined with conventional teaching laboratory equipment. In order to solve this limitation, one can exploit the fact that carbon fibers are usually supplied in bundles composed by thousands of individual fibers whose diameter is well-defined and provided by the manufacturer. For instance, the carbon fiber employed in this work was supplied in bundles composed by 12000 individual fibers with 7 μm in diameter. Therefore, one can estimate the mass density by weighting a bundle with length *L*:

$M = (N·\rho A)·L + M_0$    [2]

Where *N* is the number of individual fibers composing the bundle (number provided by the manufacturer), *ρ* is the mass density, *A* is the cross section of an individual fiber ($A = \pi r^2 = 38.5$ μm$^2$), *L* is the length of the fiber bundle and $M_0$ is the tare weight of the measurement. A better estimate of the mass density can be obtained by weighting bundles with different lengths and plotting the measured mass as a function of the filament length. To perform the *M vs. L* measurement, a fiber bundle ~ 130 mm long was employed. One of the free ends was wrapped with adhesive tape to avoid losing individual fibers during the measurement. After weighting the bundle, few millimeters of the bundle are cut with a blade. The new length of the bundle is measured with a caliber and it is weighted again. This process is repeated until obtaining several *M vs. L* data points. The mass density can be easily determined from the slope of the *M vs. L* relationship and the tare weight, discounting the tare weight.

Figure 2 shows the measured mass for a filament of 12000 carbon fibers with different lengths. The mass follows a linear trend with the filament thickness, as expected from Eq. [2]. From the slope of the *M vs. L* relationship, one can determine the mass density $\rho = 1.71 \pm 0.06$ g·cm$^{-3}$ which is in good agreement with the value provided by the manufacturer $\rho = 1.8$ g·cm$^{-3}$.

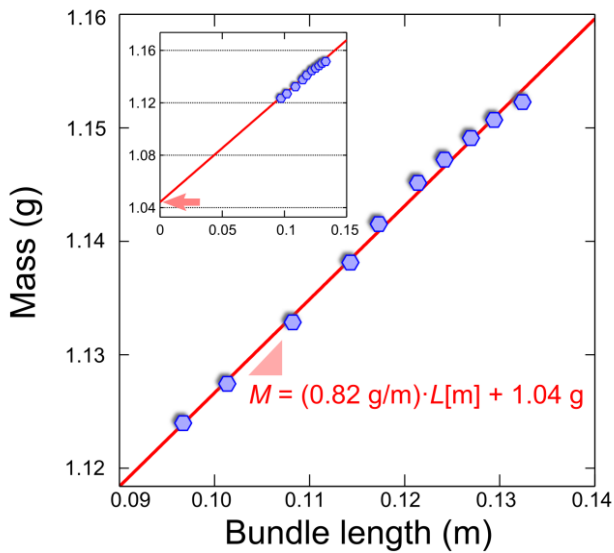

**Figure 2**: Measurement of the mass of a carbon fiber bundle composed by 12000 individual fibers. The bundle has been successively cut to obtain several pair of mass and length values. The inset indicates how the tare weight can be determined from the interception with the vertical axis.

*3.2. Young's modulus*

Young's modulus measurements on fibers are typically carried out by means of tensile and bending tests. These methods, however, are difficult to implement for micro-fibers in a teaching laboratory. Here we propose to employ a resonant method to determine the Young's modulus of micron-fibers that can be implemented in most physics teaching laboratories. The method consists on the measurement of the resonance frequency of cantilevers made from the carbon micro-fiber. For circular cross-section cantilevers, the relationship between the Young's modulus (*E*) and its resonance frequency (*f*) is given by [6]

$$E = (1-v^2)\cdot(4/3)\cdot\rho\cdot(\pi f \cdot L^2 \cdot r^{-1})^2 \quad [3]$$

Where *v* is the Poisson's ratio (*v* = 0.27 [7]), *ρ* is the mass density, *L* is the cantilever length and *r* is the fiber radius. Therefore, the Young's modulus of the carbon fibers can be obtained by simply measuring the resonance frequency of a carbon-fiber cantilever with a certain length. The measurement of the resonance frequency is carried out with the experimental setup depicted in Figure 3a. The carbon-fiber cantilever is fabricated by placing a long carbon fiber (few centimeters long) overhanging on a metal block. The fiber is fixed to the block by means of silver-loaded conductive epoxy. After curing the epoxy, the overhanging part of the fiber is cut to the desired length (typically 1 mm to 15 mm) with a blade. Then, the cantilever is driven by the electrostatic force between a metallic plate (typically a blank piece of printed circuit board connected to a signal generator) and the cantilever (grounded). The oscillation of the cantilever can be easily detected by inspecting the free end of the cantilever under an optical microscopy. Figure 3b shows an optical microscopy image of the free-end of a carbon-fiber cantilever. When the cantilever is excited at its resonance frequency, the free-end of the cantilever becomes blurred and the oscillation amplitude can be obtained directly from the image. Moreover, using stroboscopic illumination one can capture the oscillation of the carbon-fiber cantilever with a conventional camera [8, 9]. By measuring the oscillation amplitude for different excitation frequencies, one can determine the resonance spectrum of the cantilever (see Figure 3c). By fitting the resonance spectrum to a Lorentzian function, the resonance frequency and the quality factor can be determined.

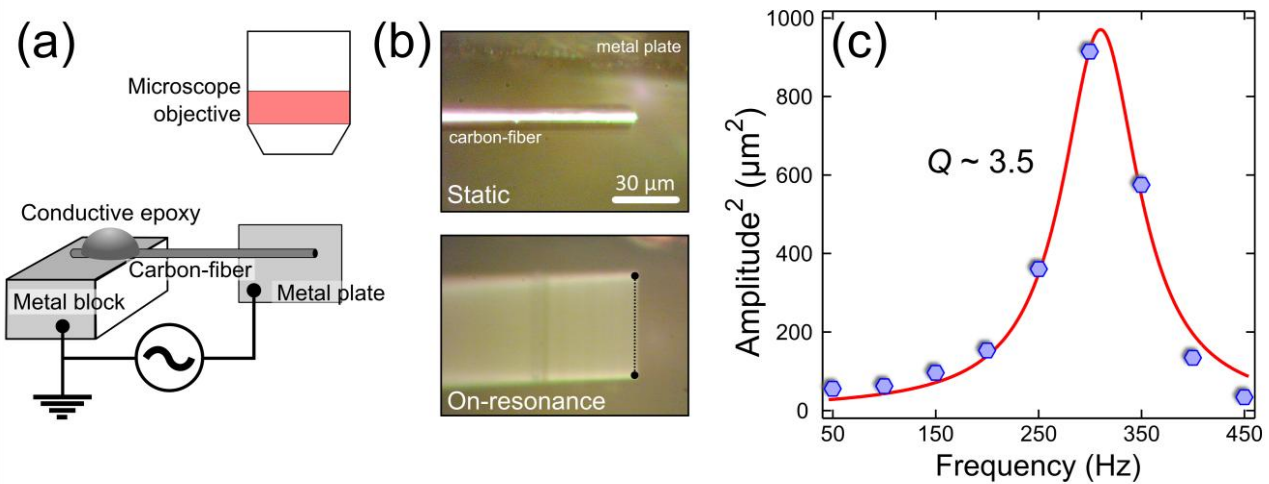

**Figure 3**: (a) Schematic diagram of the experimental setup employed to measure the resonance frequency of carbon-fiber based cantilevers. (b) Optical microscopy image of the free end of the carbon fiber cantilever in the static and on-resonance situations. (c) Resonance spectrum of a carbon fiber cantilever measured by determining the oscillation amplitude from optical microscopy images acquired while exciting the cantilever at different frequencies.

The Young's modulus of carbon micro-fibers can be determined with higher accuracy by measuring the resonance frequency on several cantilevers fabricated with different lengths. We have measured the resonance frequency of three carbon fiber cantilevers. After measuring the resonance frequency the cantilevers have been cut with a blade, the new length has been determined by means of an optical microscopy image and the new resonance frequency has been measured. This process has been carried out with each of the cantilevers to increase the statistics. Figure 4 shows the resulting measured resonance frequencies for cantilevers made from different carbon fibers (datapoints with different colors), reducing their length in steps. The Young's modulus obtained from Eq. [3] and the experimental $f$ vs. $L^{-2}$ relationship is $E = 246 \pm 8$ GPa.

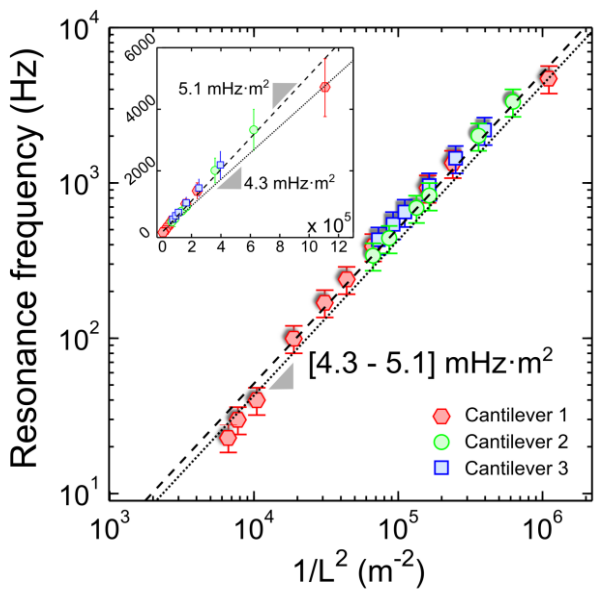

**Figure 4**: Relationship between the resonance frequency of the carbon fiber cantilevers and their length. (Inset) same relationship but represented in linear scale.

## 4. Conclusions

In summary, we presented a simple approach to characterize the electrical and mechanical properties of micro-fibers that can be carried out by physics undergraduate students in most of teaching laboratories. We applied this procedure to study carbon fibers obtaining values of the electrical resistivity, mass density and Young's modulus very close to the reference values.

## 5. Materials and methods

The carbon fibers used in this work are polyacrylonitrile (PAN), manufactured by Hercules Incorporated with part number AS4-12K. The fibers are supplied in a bundle of 12000 individual carbon fibers. The diameter of the individual fibers is 7 μm. These kind of carbon fiber bundles are typically about 10 € per meter. Table 1 summarizes the physical properties of the carbon fiber used in this work.

| Property | Value (reference) | Value (this work) |
|---|---|---|
| Electrical resistivity ($\rho_{el}$) | $1.4 \cdot 10^{-5}$ Ω·m – $1.6 \cdot 10^{-5}$ Ω·m * | $(1.3 \pm 0.3) \cdot 10^{-5}$ Ω·m |
| Mass density ($\rho$) | 1.8 gr/cm$^3$ | $1.71 \pm 0.06$ gr/cm$^3$ |
| Young's modulus ($E$) | 248 GPa | $246 \pm 8$ GPa |

**Table 1**: Physical properties of the carbon fibers employed in this work. The value with (*) has not been provided by the manufacturer and approximate values based on similar carbon fibers have been used as an estimate.

The length of the carbon fiber cantilevers and their oscillation amplitude have been determined with an optical microscope (Nikon Eclipse LV-100), previously calibrated by imaging samples of known dimensions (such as the divisions of a caliber or the diameter of the carbon fiber itself). Nonetheless, due to the dimensions of the carbon fibers one could employ a low-cost digital microscope [2] to perform the characterization presented here.

### Acknowledgements

The authors would like to acknowledge A. Lara-Quintanilla (TU Delft) for carefully proof-reading the manuscript. A.C.-G. acknowledges financial support through the FP7-Marie Curie Project PIEF-GA-2011-300802 ('STRENGTHNANO').

### Appendix

An alternative method to determine the Young's modulus of the carbon fibers consist on measuring the resonance frequency of a carbon fiber cantilever before and after loading it with test masses [10]. Before the mass load, the resonance frequency of the cantilever can be written as:

$f_0 = (1/2\pi) \cdot (k/m_{eff})^{1/2}$     [4]

---

[2] http://www.dino-lite.com/

where $k$ is the spring constant of the cantilever and $m_{eff}$ is its effective mass. After loading the cantilever with a test mass ($\Delta m$), the resonance frequency changes

$f = (1/2\pi)\cdot[k/(m_{eff} + \Delta m)]^{1/2}$    [5]

The spring constant of the cantilever can be extracted from Expressions [4] and [5] as follows:

$k = (\Delta m/4\pi^2)\,[(f_0^2 \cdot f^2) / (f_0^2 - f^2)]$    [6]

For a cantilever with circular cross section, the relationship between the spring constant and the Young's modulus is given in Ref. [6]:

$E = 2(1-v^2)(L^3/r^4)k$    [7]

Therefore, using Eq. [6] and [7] one can determine the carbon fiber Young's modulus in an alternative way.

We have measured the resonance frequency of a 1 mm long carbon fiber cantilever before and after attaching two tin microspheres (see Figure A1). The microspheres are extracted from SN62 MP218 solder paste. The test masses naturally stick to the cantilever due to the presence of a small amount of flux covering them. Nonetheless, the mass attachment procedure can be cumbersome. Once the sphericity of the particles is checked under the optical microscope, their masses are determined by measuring their diameter using the density of the bulk material (8.824 gr/cm$^3$). We have estimated that the flux increases the mass load by less than 1%.

|  | Value |
|---|---|
| Frequency before ($f$) | 4550 Hz |
| Frequency after ($f_0$) | 1390 Hz |
| Cantilever length ($L$) | 1 mm |
| Test masses ($\Delta m$) | 0.24 µgr |
| Cantilever spring constant ($k$) | 0.02 N/m |
| Young's modulus ($E$) | 231 ± 18 GPa |

**Table A1**: Values employed for the determination of the Young's modulus following the alternative approach described in the appendix.

Table A1 summarizes the values employed for the calculation of the Young's modulus. The obtained value $E = 231 \pm 18$ GPa is in very good agreement with both the reference value and the value obtained in the analysis described in the main text.

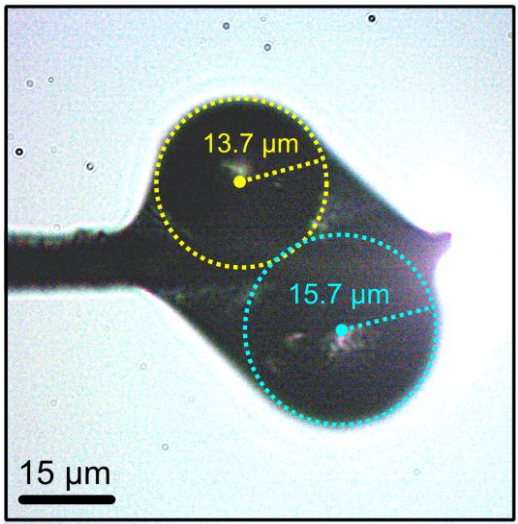

**Figure A1**: Optical microscopy image of the free end of a carbon fiber cantilever loaded with two tin microspheres.